\documentclass[prl,twocolumn,aps,amsmath,amssymb,showpacs,superscriptaddress]{revtex4-1}
\usepackage{upgreek}
\usepackage{graphicx}

\usepackage{epstopdf}

\begin{document}
\title{Buckled Diamond-like Carbon Nanomechanical Resonators }

\author{Matti Tomi}
\affiliation{Low Temperature Laboratory, Deptartment of Physics, Aalto University School of Science, FI-00076, Finland}
\author{Andreas Isacsson}
\affiliation{Department of Applied Physics, Chalmers University of Technology, SE-412 96 G{\"o}teborg, Sweden}
\author{Mika Oksanen}
\affiliation{Low Temperature Laboratory, Deptartment of Physics, Aalto University School of Science, FI-00076, Finland}
\author{Dmitry Lyashenko}
\affiliation{Low Temperature Laboratory, Deptartment of Physics, Aalto University School of Science, FI-00076, Finland}
\author{ Jukka-Pekka Kaikkonen}
\affiliation{Low Temperature Laboratory, Deptartment of Physics, Aalto University School of Science, FI-00076, Finland}
\author{Sanna Tervakangas}
\affiliation{DIARC-Technology Oy, Kattilalaaksontie 1, FI-02330 Espoo, Finland}
\author{Jukka Kolehmainen}
\affiliation{DIARC-Technology Oy, Kattilalaaksontie 1, FI-02330 Espoo, Finland}
\author{Pertti J. Hakonen} \email[Corresponding author: pertti.hakonen@aalto.fi]{} 
\affiliation{Low Temperature Laboratory, Deptartment of Physics, Aalto University School of Science, FI-00076, Finland}

\begin{abstract}

We have developed capacitively-transduced nanomechanical resonators using sp$^2$-rich diamond-like carbon (DLC) thin films as conducting membranes. The electrically conducting DLC films were grown by physical vapor deposition at a temperature of $500{\,\,}^\circ$C. Characterizing the resonant response, we find a larger than expected frequency tuning that we attribute to the membrane being buckled upwards, away from the bottom electrode. The possibility of using buckled resonators to increase frequency tuning can be of advantage in rf applications such as tunable GHz filters and voltage-controlled oscillators.

\end{abstract}


\maketitle

\section{Introduction}
Several device applications based on nanoelectromechanical (NEM) resonators benefit from maximizing the resonator area in order to achieve good performance. To preserve the benefits coming from the small linear size, such as high operating frequency and sensitivity, this implies that finding materials with low mass density, large mechanical stiffness and good electrical conductivity is important. In addition, the material should be suitable for the fabrication of suspended membrane or plate-like geometries. For these reasons, several recent devices have employed single- or few-layer graphene~\cite{Bunch_2007, Hone_2009, Desmukh_2010, Oksanen_2011, Eichler_2011}, MoS$_2$~\cite{Feng_2013, Castellanos_Gomez_2013} or graphene-coated SiN~\cite{Parpia_2013}. In the latter case, graphene was added to enable electrical readout.

Here we report on the fabrication and characterization of plate-like NEM resonators made from diamond-like amorphous carbon (DLC)~\cite{Robertson_2002} with electrical transduction (see Fig.~\ref{fig:device}). Due to the strong sp$^3$ bond, DLC shares the beneficial mechanical properties of graphene and diamond, such as low mass and high stiffness. However, as DLC typically conducts poorly due to its low sp$^2$ content, fully electrostatic transduction has not been previously reported. Although electrostatic actuation has been achieved for DLC MEMS/NEMS, the devices have been reliant on metal coatings and/or optical
readout~\cite{DLC_elres1, DLC_res1, DLC_res2, Chua2004, Sekaric2002, Zalalutdinov2011, Hirono2002, Seshan2015}, limiting the technological application potential of DLC in MEMS/NEMS~\cite{Liao2011,DLCmems}. In our resonators, DLC with a high sp$^2$ content was used to overcome this problem, enabling a direct capacitive electrical readout.

Using DLC, rather than graphene for instance, has also other advantages. Monolayer graphene, although very rigid, will easily strain so much due to electrostatic forces that the capacitor plates snap together at what is known as the pull-in point. Although any device will experience pull-in at some level, for monolayer graphene it occurs at a relatively low electric field. While employing multilayer graphene allows the use of higher electric fields, it is more difficult to manufacture in bulk. Using DLC averts this problem, as it can be easily produced in large quantities by means of high-pressure high-temperature (HPHT) synthesis, chemical vapour deposition (CVD) or filtered cathodic vacuum arc (FCVA) techniques. Moreover, during the growth process it is possible to retain control over the composition and structure of the material, including its mechanical, electrical and optical properties~\cite{Williams2011}.

A characteristic feature of NEM resonators is the tuning of the resonant frequency while changing the dc bias voltage. Having a large frequency tuning in conjunction with a high operating frequency is of technological importance in rf applications. The tuning in NEM resonators has two main sources; partly it arises from the electrostatic spring softening as the resonator is deflected in the nonlinear electrostatic field, leading to a decrease in frequency with increasing dc voltage. The second source is the geometric nonlinearity which occurs due to a change in the length of the resonator, typically resulting in an increased frequency caused by the induced tension~\cite{Lifshitz_review}. While reaching high frequencies can be done by increasing the resonator thickness, this usually comes at the expense of smaller frequency tuning within the accessible bias range. For applications reliant on tunable resonators in the GHz regime, such as filters and voltage-controlled oscillators (VCO)\cite{Chen_2013}, raising the base frequency while maintaining a large tuning is desirable.

We find in our devices a considerably larger tuning downward in frequency than can be attributed to electrostatic spring softening alone. As the DLC films have built-in compressive stress, the resonators display Euler buckling instabilities upon fabrication. This affects the frequency tuning~\cite{Nayfeh_buckled1, Nayfeh_buckled2, Parpia_buckled}. The tuning curves for the DLC resonators can be fitted, assuming that the suspended DLC is buckled upwards, $i.e.$ away from the bottom electrode. The frequency shift here is dominated by the geometric nonlinearity, which in this case causes a decrease in frequency with increasing dc voltage. Hence, by using buckled DLC resonators the tuning range can be extended at moderate bias voltages even for thicker structures.

\begin{figure}
\includegraphics*[width=0.8\linewidth]{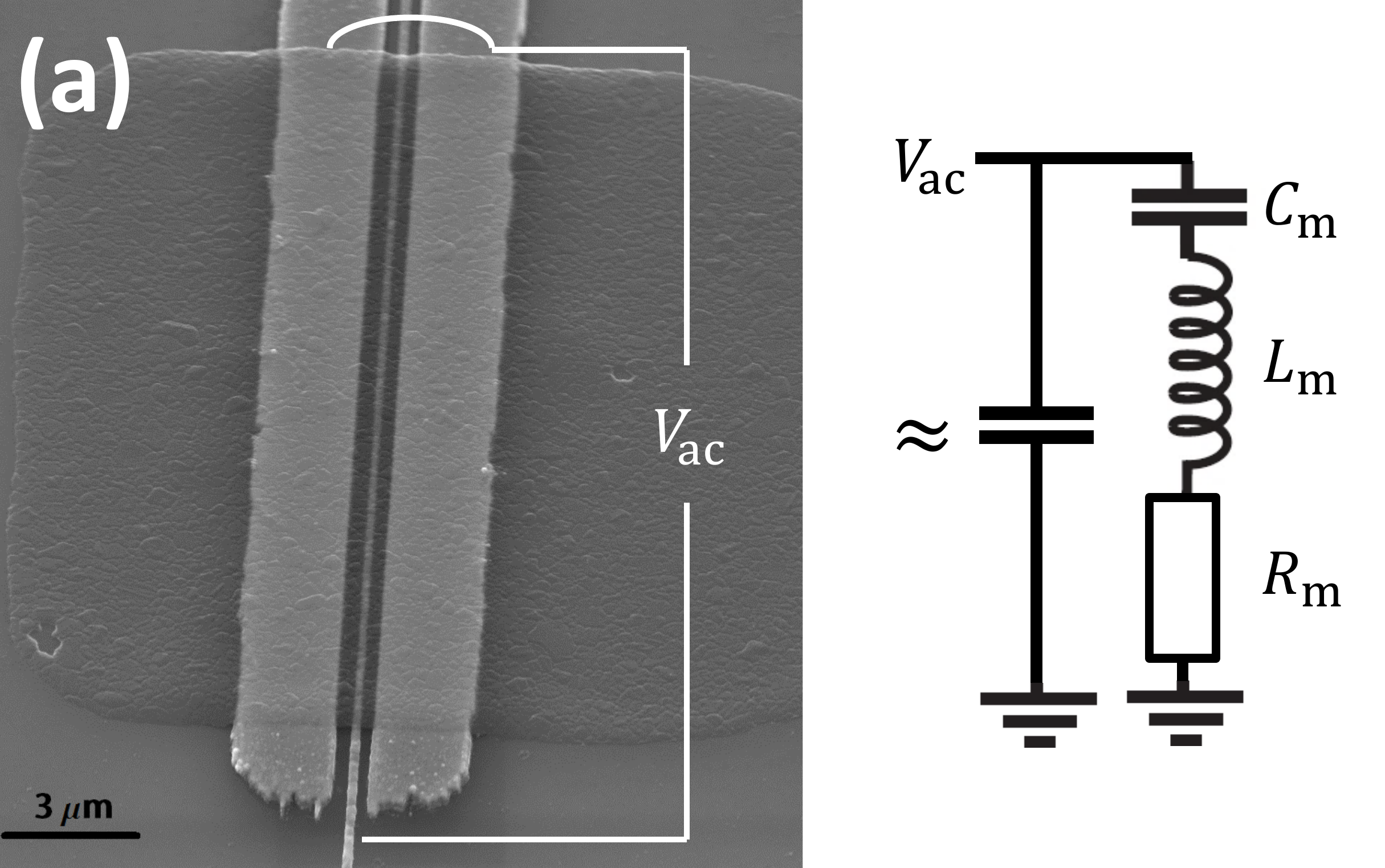}
\includegraphics*[width=0.8\linewidth]{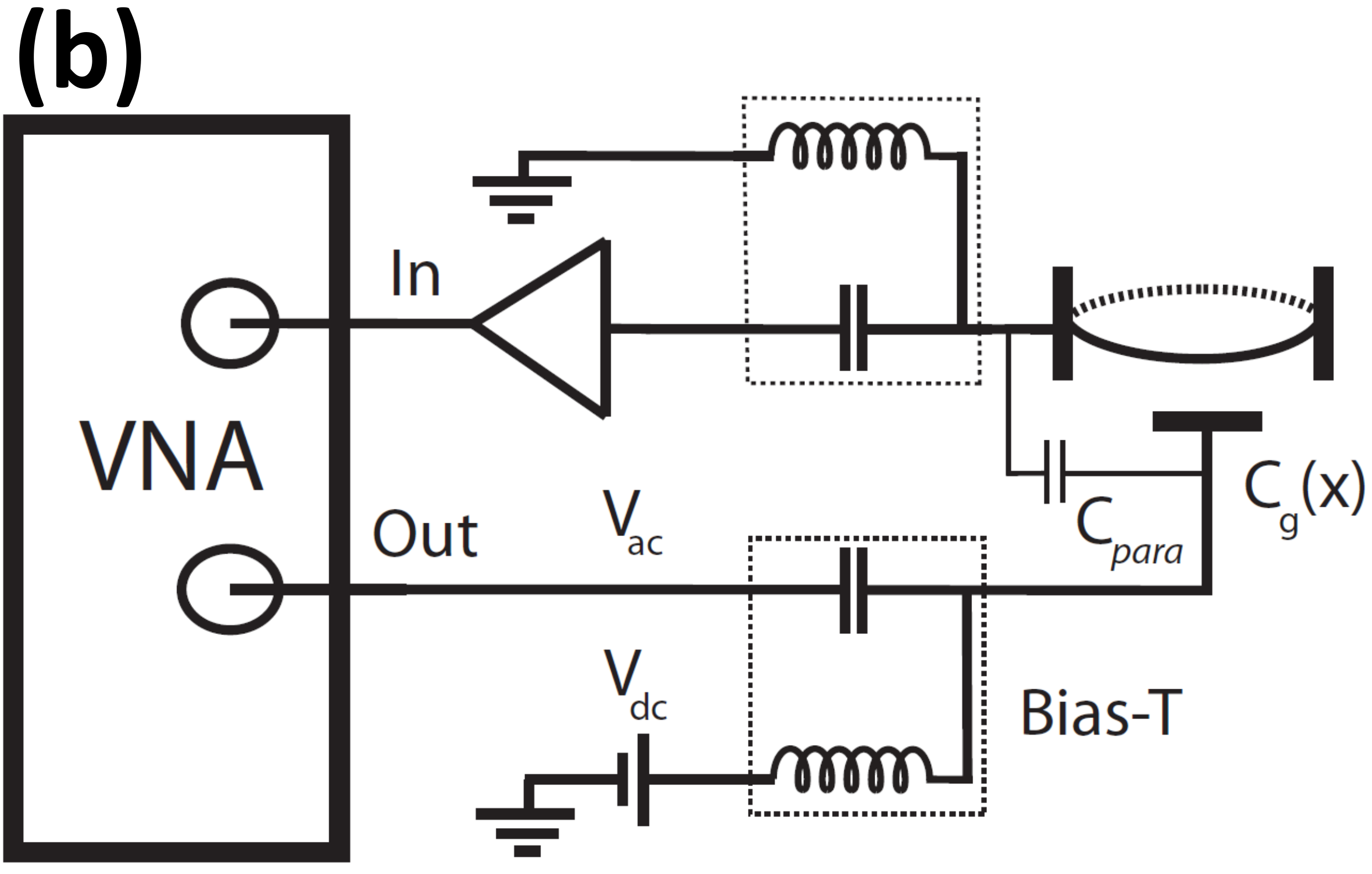}
\caption{ (a) SEM image of a DLC  resonator with a suspended area of 15$\times$1~${\upmu}$m$^2$. Our device layout consists of the DLC supported by a contact fork, a distance of $d_0=205$~nm above a bottom electrode. Applying an ac signal superimposed on a dc voltage between the electrodes, the electrical response is that of an RLC series resonator with dc-bias-dependent equivalent circuit elements $R_{\rm m}$, $L_{\rm m}$, and $C_{\rm m}$. The LC resonance frequency is equal to the mechanical resonance. (b): Schematic of the measurement setup: a vector network analyzer (VNA) is employed for direct transmission measurements through the capacitive resonator $C_{\rm g}(x)$ under study. The device is biased via two bias-T components which provide a voltage $V_\mathrm{dc}$ across the sample.}
\label{fig:device}
\end{figure}

\section{Fabrication}
The 20~nm thick conductive DLC films were grown by physical vapour deposition (PVD) techniques\cite{pvd} at a temperature of 500~${}^\circ$C. The thickness was determined using a profilometer. To facilitate chemical release from the substrate, the DLC films were grown on a Si substrate with a 100~nm sacrificial layer of Al or Cu. This resulted in films with a square resistance at room temperature of about 370 $\Omega$/sq  (2.2 k$\Omega$/sq at 4.2~K), which corresponds to a resistivity of $7.4\times 10^{-4}$~$\Omega\cdot$cm. According to Raman spectra, the sp$^3$ content in our films is around 10~\% and, consequently, they can be classified as "sp$^2$-rich tetrahedral amorphous carbon (ta-C) films"~\cite{Robertson_2002}. Atomic force microscope (AFM) measurements of the surface topology of the films revealed an rms surface roughness of $\sim$5~nm\dag \footnote{\dag See supplementary information at the of the manuscript}. The residual compressive stress in the as-deposited film was determined to be approximately 2~GPa\cite{prestress}.

The fork-shaped metallic electrode structures seen in Fig.~\ref{fig:device}a were fabricated in two steps. First, we deposited thick 255~nm Au electrodes on a 100~mm high-purity Si/SiO${}_{2}$ wafer using optical lithography and electron-beam evaporation. These support electrodes have a 1.0~${\upmu}$m wide and 50~${\upmu}$m long trench, on top of which the DLC film was later transferred. The wafer was diced using a diamond-bladed saw into 5~mm $\times$ 5~mm chips with 16 electrode structures each. In a second step, a thin bottom electrode with a width of 400~nm to 600~nm and a thickness of 50~nm, was deposited in the middle of each trench using electron-beam lithography and thermal evaporation. Bias voltages up to 100~V dc could be applied across the gap structures under ultra-high vacuum conditions at liquid helium temperatures.

To suspend the DLC films they were first supported from the top side by spinning a layer of poly(methyl methacrylate) (PMMA) resist on top of the films. The sacrificial layer was then removed by wet etching in either 10 \% HCl (for Al) or FeCl${}_{3}$ (for Cu). The DLC/PMMA membrane was rinsed in deionized water and deposited on the target substrate, directly on top of the support electrode. As a final step, the PMMA layer was removed by baking the chip in a hydrogen atmosphere (5~\% H${}_{2}$ in Ar) at 375 ${}^\circ$C for several hours. Using this procedure, we obtained suspended DLC films with a length of 1~${\upmu}$m, a width up to 50 ${\upmu}$m, and a distance to the bottom electrode which was typically around 200 nm (Fig.~\ref{fig:device}).

\section{Results}
Measurements were carried out at a temperature of 4.2~K under a residual gas pressure below 10${}^{-5}$~mbar. A dc voltage was applied between the DLC film and the bottom electrode, and resonator motion was actuated by an rf signal from a vector network analyzer (VNA). Transmission through the device was measured with the VNA as a function of frequency and dc voltage as illustrated in Fig.~\ref{fig:device}b. The typical rf power in the measurements ranged from $-40$~dBm to $-30$~dBm.

\begin{figure}[t]
\includegraphics[width=0.8\linewidth]{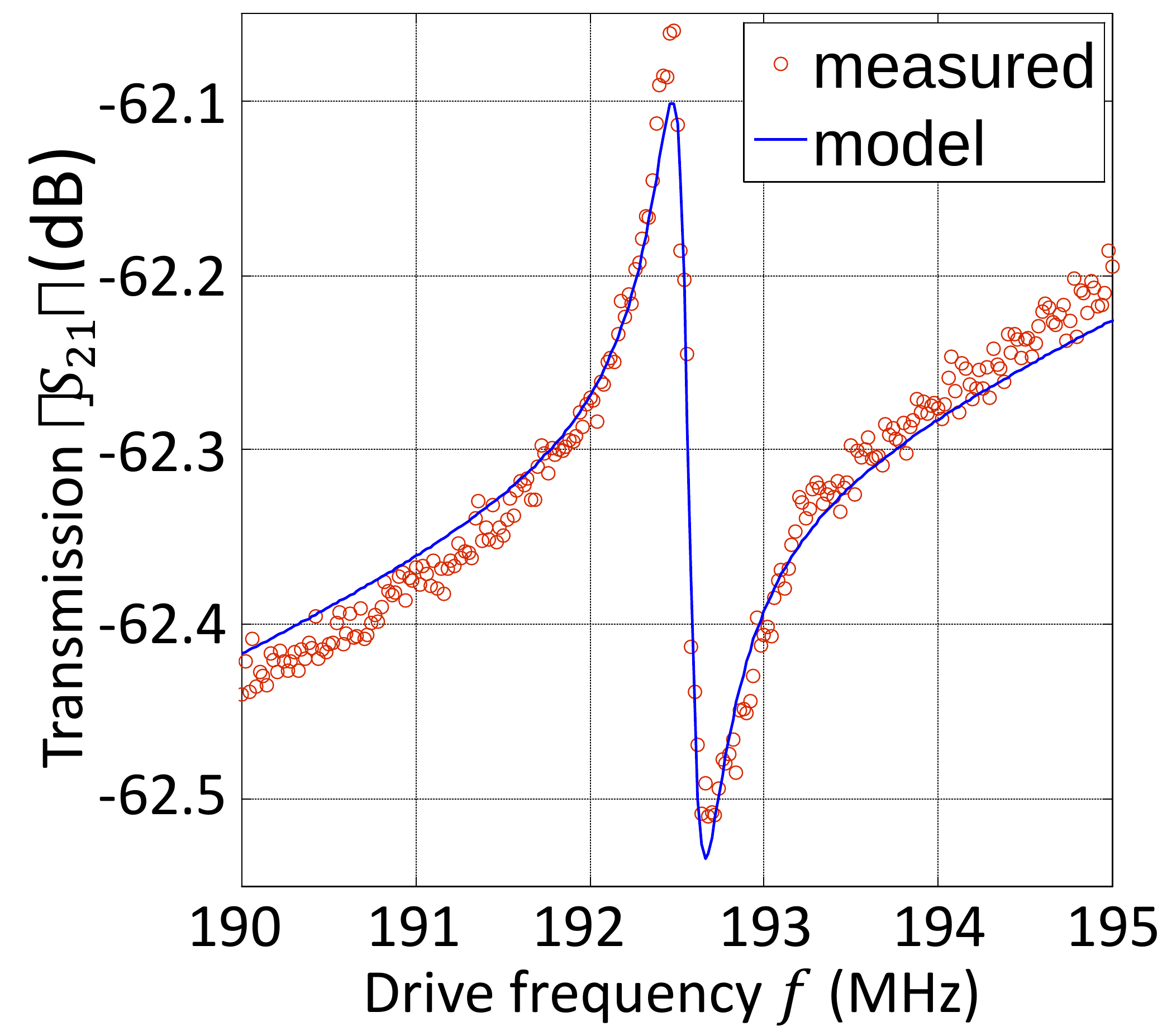}
\caption{Transmission through the DLC resonator at dc bias $V_\mathrm{dc}$ = 10~V compared with electrical modeling using the equivalent RLC circuit in Fig.~\ref{fig:device}(a). The baseline is given by parasitic capacitances and the modelled transmission magnitude (blue) is calculated using a parasitic capacitance of $C_\mathrm{para}=6.3$~fF across the device under study. Other parameters are: $Q = 990$, $L = 1$~${\upmu}$m, $t = 20$ nm, $d_0 = 205$~nm, $W=4.3\,\upmu$m, $\rho = 2000$ kg/m$^3$, $P = -40$~dBm.}
\label{fig:fit}
\end{figure}

\subsection{Measurement results}
\begin{figure}[t]
\includegraphics*[width=0.9\linewidth]{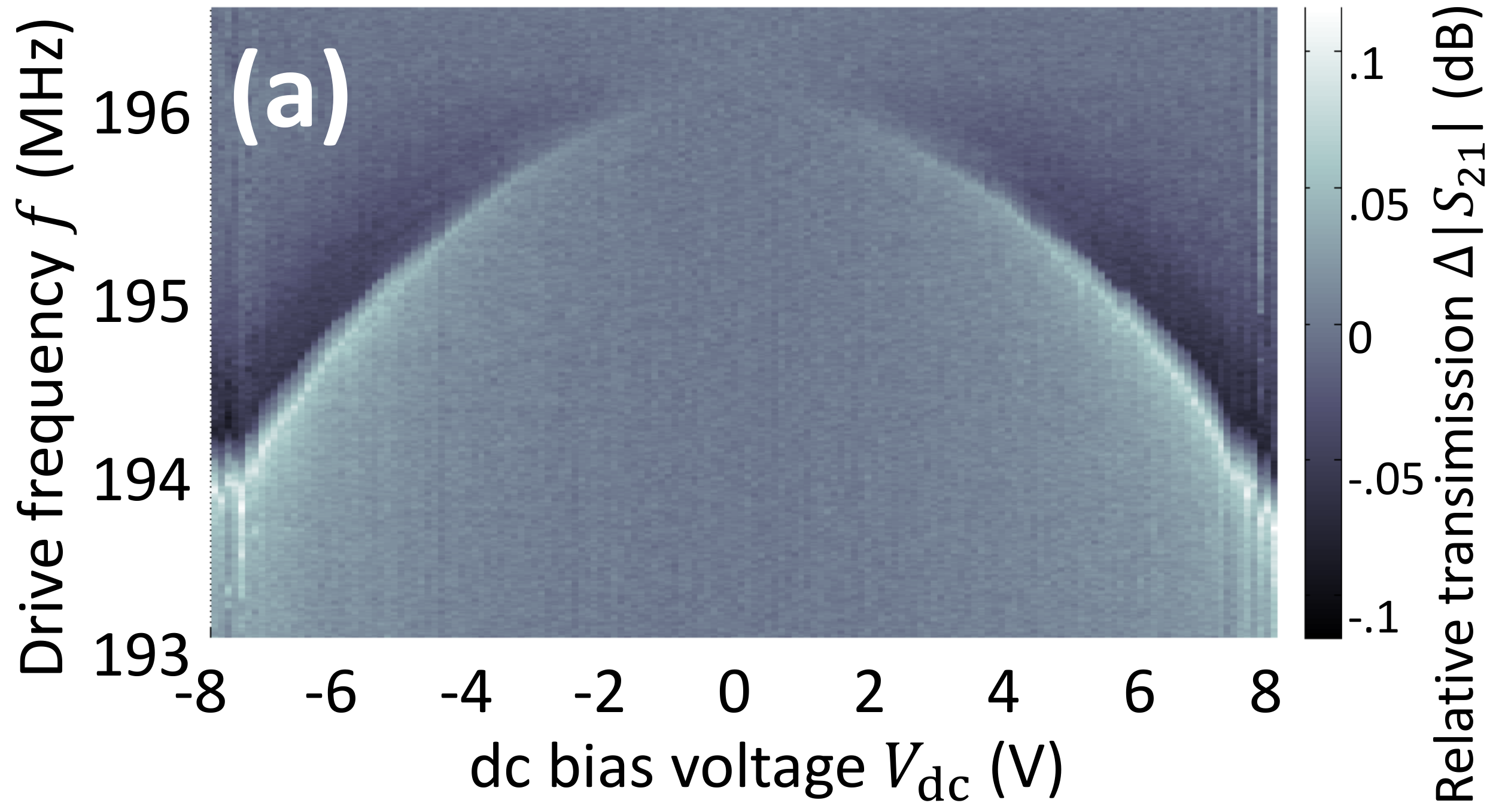}
\includegraphics*[width=0.9\linewidth]{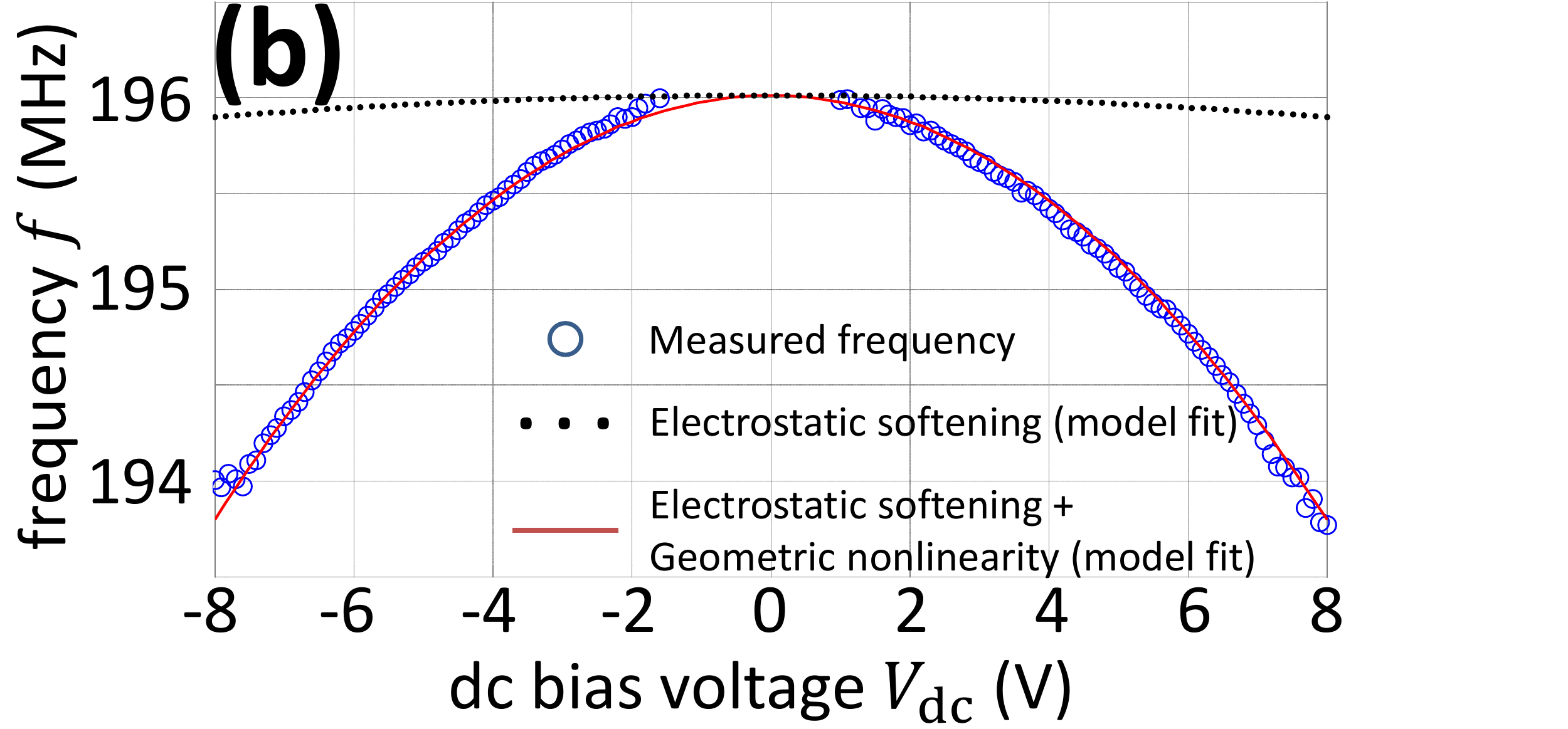}
\includegraphics*[width=0.9\linewidth]{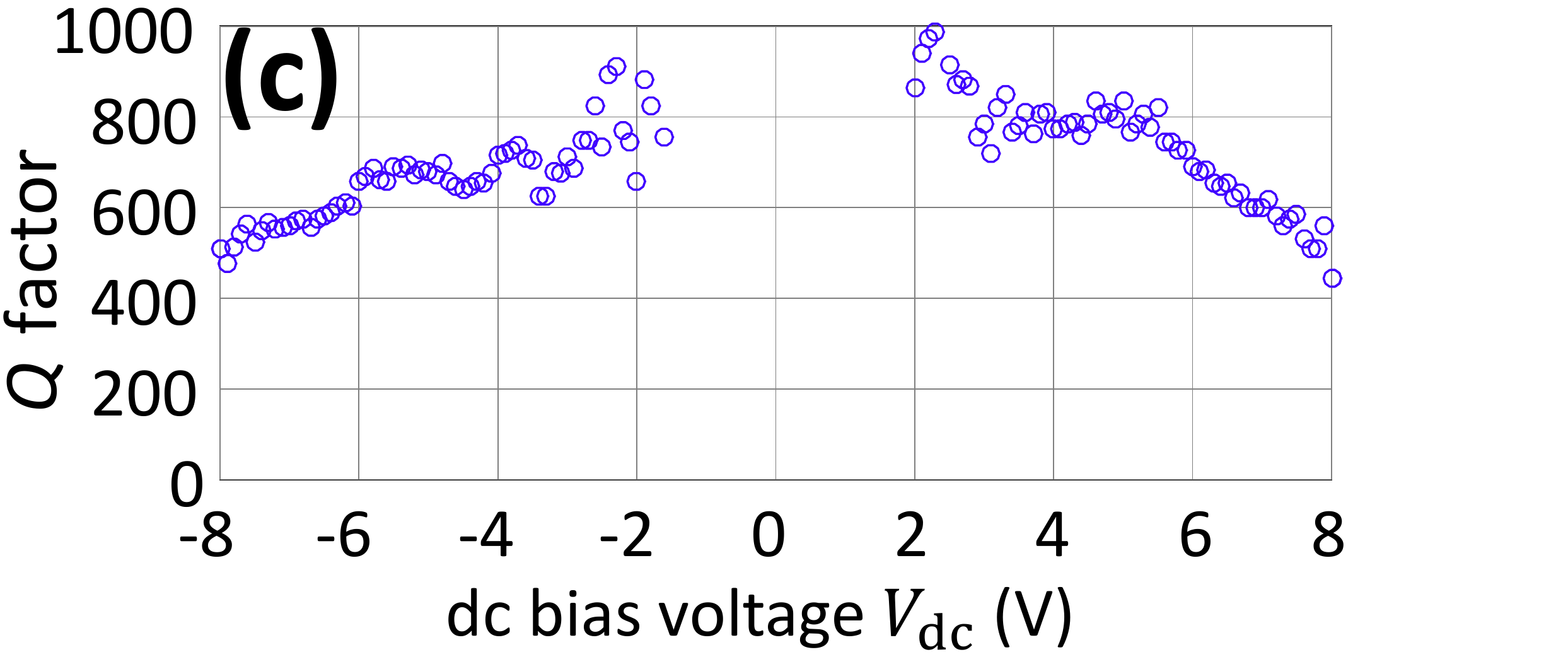}
\caption{(a) Relative transmission as a function of dc voltage and frequency at a constant ac power $P=-30$ dBm. (b) Model fit to the frequency tuning. Blue circles are measured points. The dotted black line shows the contribution to the tuning from the electrostatic softening (assuming no buckling). The solid red line is the theoretical fit assuming an upward-buckled configuration. The fits were obtained using a thickness $t=20$~nm, suspension height $d=205$~nm, mass density $\rho=2000$~kg/m$^3$ and Young's modulus $E=160$~GPa. (c) Quality factor as a function of dc voltage at a constant ac power $P=-30$ dBm.}
\label{fig:frequency}
\end{figure}

The measured transmission at a given dc bias shows a resonant feature as the one in Fig.~\ref{fig:fit}. This is consistent with a capacitively transduced resonator which can be modeled with an electrical equivalent circuit consisting of a resistor $R_{\rm m}$, a capacitor $C_{\rm m}$ and an inductor $L_{\rm m}$ in series\cite{Truitt_2007, Eriksson_2008} (see Fig.~\ref{fig:device}a), along with a parallel capacitance representing the regular current path. The measured line shape, together with the characteristic dc voltage tuning of the resonant frequency (see Fig.~\ref{fig:frequency}) is further a reliable identification of the resonance as being of mechanical origin. The resonance frequency of a 1~${\upmu}$m long and 20~nm thick DLC membrane was found to be 196~MHz at a low dc voltage (see Fig.~\ref{fig:frequency}a), with a frequency tunability of $\sim$2~\% by changing the dc voltage up to $\pm$10~V.

\subsection{Resonator characterization}
In order to characterize the resonator, the motional RLC impedances must be related to their mechanical counterparts. For a mechanical resonator with frequency $f=\Omega/2\pi$, the motional circuit elements are related to the mechanical parameters as $R_{\rm m}={\Omega m}/{Q{\eta }^2}$, $C_{\rm m}={{\eta }^2}/{m \Omega^2}$, $L_{\rm m}={m}/{{\eta }^2}$, where $m$ is the mass of the resonator and $Q$ is the mechanical quality factor. The parameter $\eta$ is the the effective electromechanical transduction factor, defined as
\begin{equation}
\eta =V_\mathrm{dc}\frac{\partial C}{\partial z}\approx \alpha V_\mathrm{dc}\frac{\varepsilon WL}{d_0^2},
\label{eq:eta}
\end{equation}
where $W$, $L$ and $d_0$ are the width, length and plate gap of the resonator, respectively; the factor $\alpha\lesssim 1$ has been added to account for the vibration mode shape and deviations from the parallel-plate capacitor model. The parameter $\alpha$ is sometimes also included by introducing an effective mass. As the resonating width $W$ is not known, we use here $\alpha=1$ and fit $W$.

Using the above transduction model, we have calculated the electrical transmission response of our DLC resonator. The result is displayed in Fig.~\ref{fig:fit} at V${}_\mathrm{dc}$ = 10~V. The simulation was done for a resonating width $W = 4.3$~${\upmu}$m, with all the other employed parameters found in the caption of Fig.~\ref{fig:fit}. According to the transduction model, the mechanical equivalent resistance is $R_{\rm m} \sim 2.6$~M$\Omega$, which is much larger than the electrical resistance of the DLC strip $R_\mathrm{DLC} \sim 500$~$\Omega$ or the impedance of the parasitic capacitance $Z_\mathrm{para} \sim 130$~k$\Omega$\dag. The width $W\approx 4.3$~$\upmu$m used in the fit reflects that in the measured devices the actuation factor was reduced by a factor of 1--10 as compared to what one would expect from Eq.~(\ref{eq:eta}) assuming $\alpha=1$ and the entire width of the suspended DLC sheet vibrating. We attribute this reduction to the fundamental mode splitting into several separate resonances as a consequence of imperfect boundary conditions at the gold--DLC interface\dag. By improving the surface smoothness of the electrodes, larger actuation factors should be achievable.

From the fitting we also obtain the quality factor Q. At low drive powers ($P=-40$~dBm) and small bias voltages (dc bias 2--5 V) we find $Q\gtrsim 1400$. As the dc bias is increased, the $Q$ factor decreases. At the 10~V bias shown in Fig.~\ref{fig:fit}, the $Q$ factor has decreased to $Q\approx 1000$ at $P=-40$~dBm. This is consistent with increased ohmic dissipation due to the induced displacement currents in the DLC film\cite{Oksanen_2011}.

We also noticed a dependence of the $Q$ factor on the ac drive power. At a higher drive power $P=-30$~dBm, the quality factor ranged from $Q\sim 1000$ at low bias, down to $Q\sim 500$ at  V${}_\mathrm{dc}$ = $\pm 8$~V (Fig.~\ref{fig:frequency}c). For a purely linear resonator system, the $Q$ factor remains unchanged with increased drive power. The effect leading to the broadening of the resonance and reduction of $Q$ may be related to nonlinear dissipation\cite{Bachtold_2011}.

\subsection{Frequency tuning}
We conclude the measurement section by considering the tuning of the resonant frequency with dc voltage. The downward tuning with increased bias (see Fig.~\ref{fig:frequency}a) is suggestive of electrostatic softening being the dominating contributor~\cite{MQT}. However, as the black dotted curve in Fig.~\ref{fig:frequency}b indicates, the predicted frequency tuning due to this mechanism cannot account for the measured tuning. An alternative explanation is that the large built-in compressive stress in the DLC is partly released when it is suspended, leading to Euler buckling.

Assuming that the buckling is upwards, away from the bottom electrode, the observed tuning of the resonance frequency can be accurately reproduced within the parameter range appropriate for our DLC resonator. The calculated\dag frequency tuning is shown in Fig.~\ref{fig:frequency}b as the red solid line. The buckled configuration appears to be quite favorable for frequency tuning, and based on our theoretical model\dag we estimate that a 20 \% tunability can be achieved by increasing the voltage up to about 30~V.

\section{Conclusions}
We have demonstrated that conductive DLC ($\rho \sim 10^{-3}\,\,\Omega\cdot$cm) can serve as a material platform for nanoelectromechanical resonators with capacitive transduction. This complements previously used materials, such as graphene, MoS$_2$, SiN/graphene and others, for applications based on membrane-like NEM resonators. As a material, DLC shares some of the mechanical properties of graphene (low mass, high stiffness), with the added benefit of established methods for bulk production.

For the DLC resonators presented here, the built-in compressive stress further led to significantly enhanced frequency tunability due to the devices operating in the Euler-buckled regime. This suggests that for devices where a large frequency tuning is desirable, such as frequency-tunable filters and VCOs, buckled NEM resonators could be used. However, the frequency curve is highly sensitive to the precise value of the built-in prestress, and reproducibility in large scale production may pose a challenge.

\section*{Acknowledgements}
This work was supported by the
European Union Seventh Framework Programme under the grant agreement no. 246026. The work was further supported by the Academy of Finland (contract no. 250280, LTQ CoE). Our work benefited from the use of the Aalto University Low Temperature Laboratory infrastructure. MO is grateful to V\"{a}is\"{a}l\"{a} Foundation of the Finnish Academy of Science and Letters for a scholarship. AI also acknowledges the hospitality of Aalto University.

\dag See the supplementary information at the end of the manuscript.
\newpage
\newcommand{\RR}{\right}
\newcommand{\LL}{\left}
\newcommand{\m}{\mathrm}
\section{Online supplementary material to Buckled Diamond-like Carbon Nanomechanical Resonators}

{\bf These supplementary notes to the paper Buckled Diamond-like Carbon Nanomechanical Resonators are organized as follows: First, we present and discuss some details of our measurement data for the diamond-like carbon (DLC) resonators. This is followed by a section where we highlight the importance of a good mechanical contact between DLC and the support structure; a feature that we believe is the origin of the smallness of the measured effective resonator width $W$. We then present measurements on surface roughness and discuss its relation to the material parameters used to fit the measured data to the model. We end by giving a detailed account for the derivation of the equations describing the eigenspectrum of buckled beams under electrostatic load.}

\section*{Transmission of the motional branch}
\begin{figure}[t]
\centering
\includegraphics[width=0.9\linewidth]{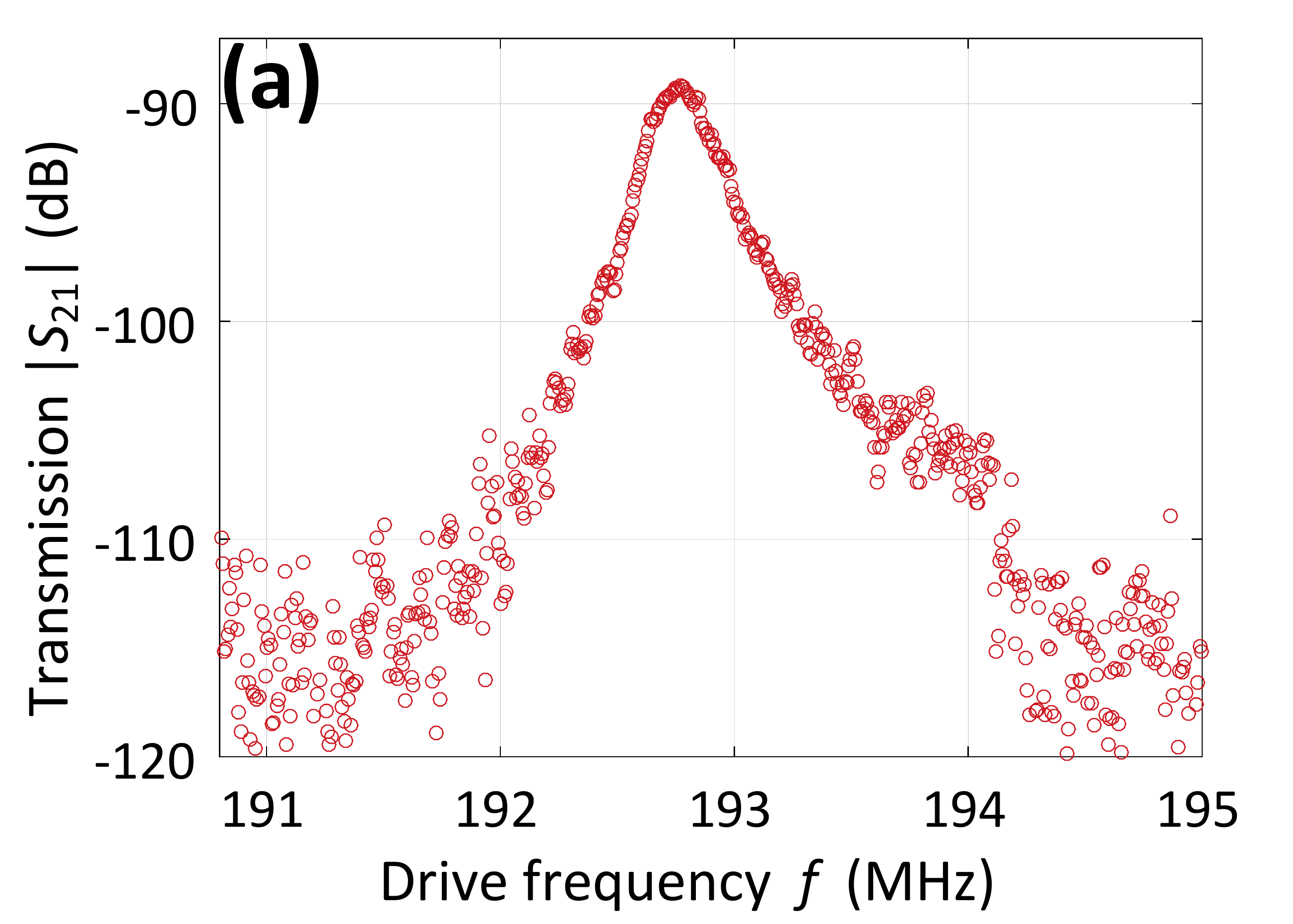}
\includegraphics[width=0.9\linewidth]{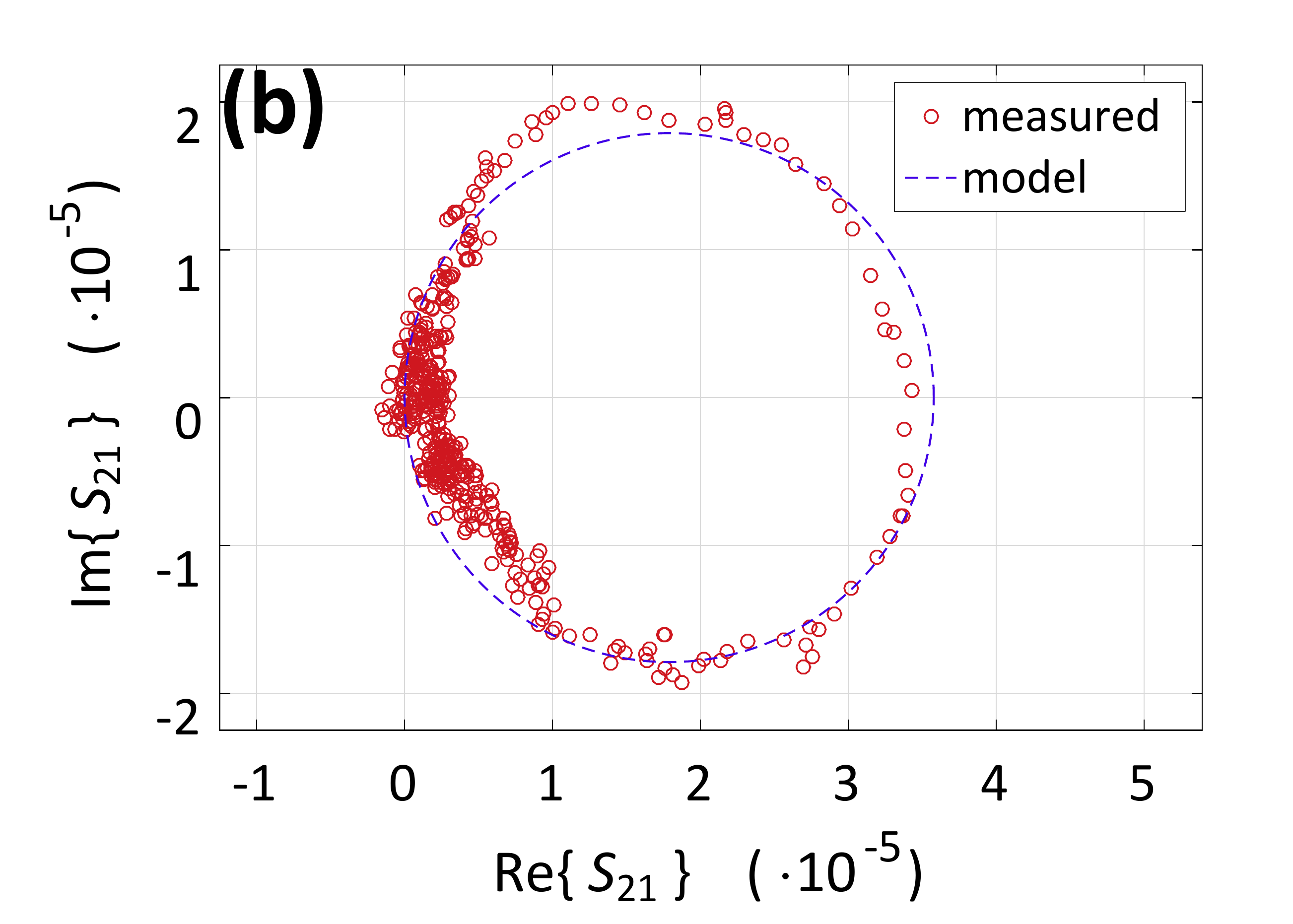}
\caption{(a) Transmission through the DLC resonator at dc bias $V_\mathrm{dc}=-10$~V and $P=-40$~dBm with parasitic components subtracted. (b) Nyquist plot of the same data in the complex plane. The dashed blue line shows a circular fit given by the (motional) RLC model.\label{fig:transmission}}
\end{figure}

Transmission through the DLC resonator device was measured with a vector network analyzer (VNA), which enabled recording both magnitude and phase information. As the impedance of the parasitic capacitance is smaller than the impedance of the motional RLC resonator, the resonance lineshape consists of an upward peak, followed immediately by a downward peak. If we subtract the vector background that arises due to parasitics, we are left with a transmission spectrum for the motional RLC branch only (Fig.~\ref{fig:transmission}a). The magnitude of this motional RLC resonance takes on the conventional Lorentzian lineshape, limited by noise from the VNA when moving far away from resonance.

To visualize the complex form of the transmission, we can plot its real and imaginary parts in the complex plane while sweeping frequency near the resonance (Fig.~\ref{fig:transmission}b). This representation, known as the Nyquist plot, should form a circular or an elliptical response depending on whether the resonance is linear or nonlinear, respectively. From Fig.~\ref{fig:transmission}b we can deduce that no prominent nonlinear effects are present in the DLC resonator at low drive powers.

\section*{Imperfect boundary conditions}
In our measurements we have found that sometimes the boundary conditions of the resonator are not well defined. This is because of the weak adhesion of DLC to the supporting Au electrodes. The compressive stress in the DLC film may partially detach it from the Au surface, which leads to ambiguous boundary conditions. These, in turn, lead to the localization of the resonance modes. Fig.~\ref{fig:device} shows scanning electron microscope (SEM) and optical images of one particular device with very poor attachment, and where several resonances were detected. The effect of the imperfect boundary conditions is thus that the actual width $W$ of the vibrating region can be considerably smaller than the width of the entire DLC sheet. Also, in the better-contacted devices, small irregularities in the clamping can be sufficient to cause splitting of the modes.

\begin{figure}[t]
\centering
\includegraphics[width=0.9\linewidth]{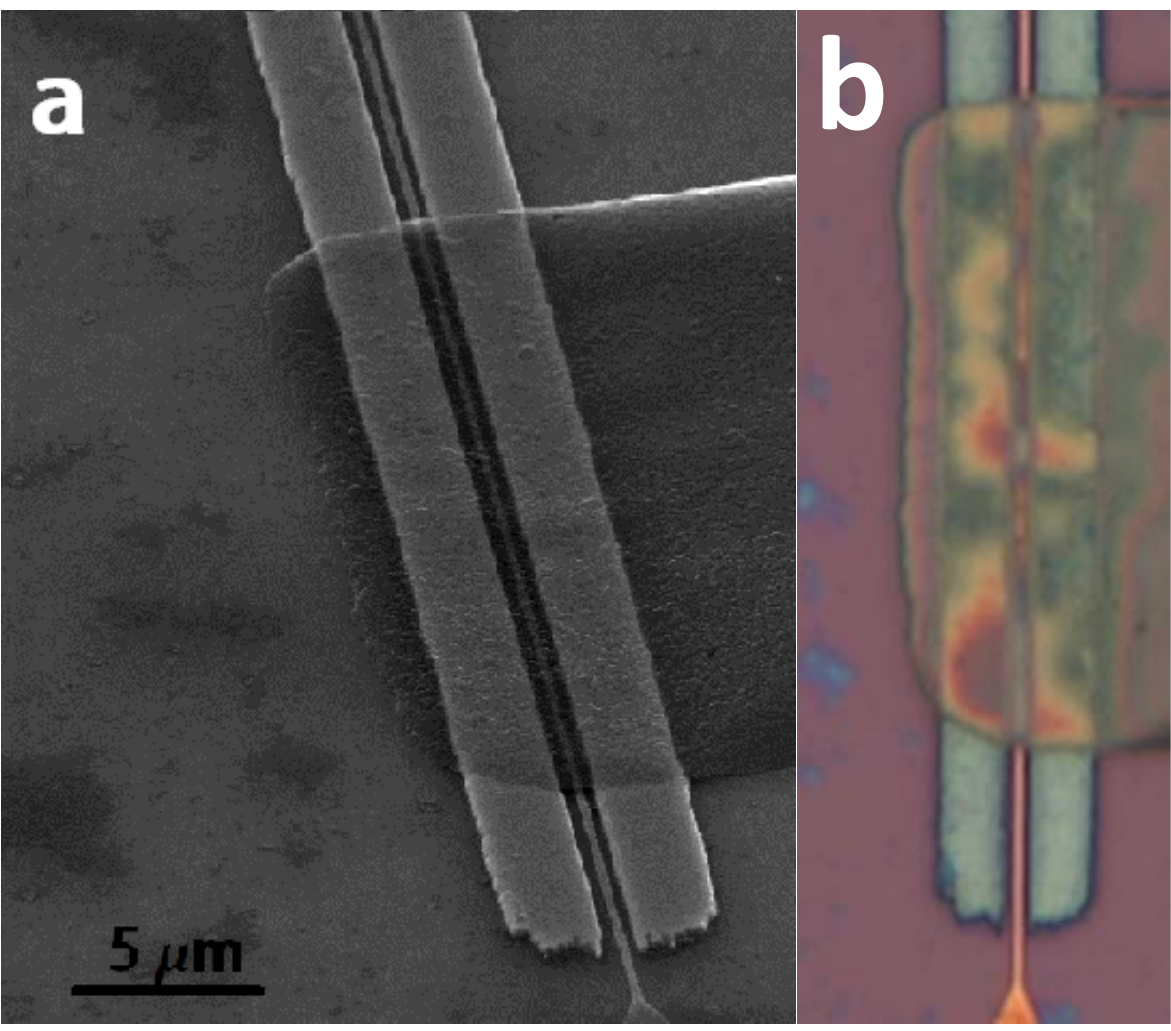}
\caption{(a) SEM image of a DLC resonator with imperfect boundary conditions. (b) Optical image of the device, which shows the irregularities in the clamping. The irregularities split the uniform oscillations into several separate modes. This reduces the width $W$ of the resonating region.\label{fig:device}}
\end{figure}

\begin{figure}
\begin{center}
\includegraphics*[width=0.9\linewidth]{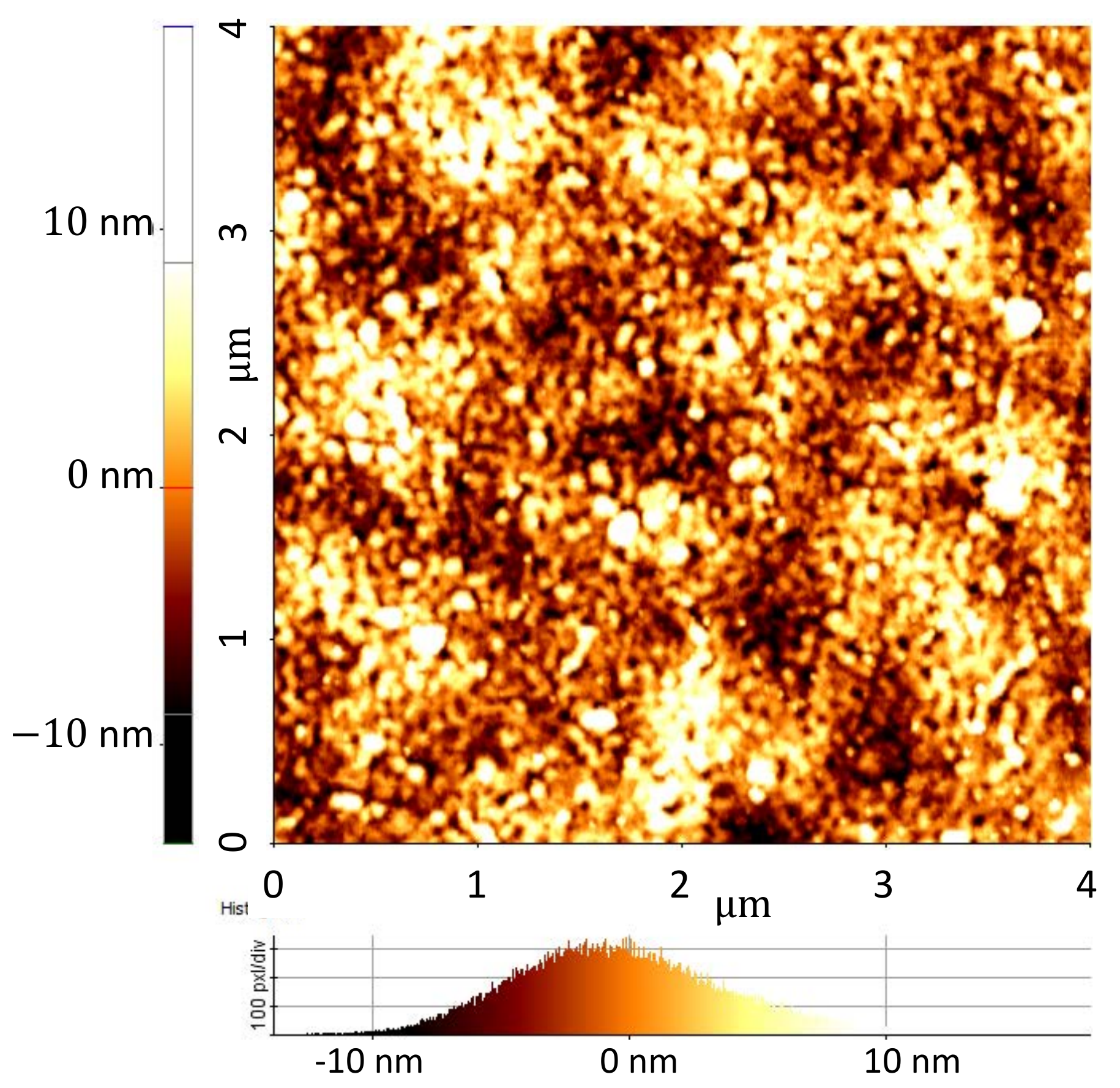}
\caption{(top) False colour AFM image of the DLC film. (bottom) Histogram over the surface height distribution.  \label{fig:AFMfig}}
\end{center}
\end{figure}
\section*{Fitting of the frequency tuning and surface roughness}
While some of the device parameters are known to good accuracy, others, such as the thickness $t$, are harder to quantify precisely. For nanoresonators it is known that the effective elastic properties change as the surface-to-volume ratio increases\cite{Delft}. Hence, for a resonator made of a material with Young's modulus $E$, an effective modulus $E^*\neq E$ may be needed.

The DLC films used here nominally have a thickness of $t\approx 20$~nm, measured using a DektakXT profilometer. However, the degree of surface roughness arising from the fabrication process is relatively large, with a full width at half maximum (FWHM) on the order of 10~nm as characterized by atomic force microscope (AFM) measurements (see Fig.~\ref{fig:AFMfig}). This suggests that the dominating surface effects in our case stem from the surface roughness.

A common model for including the effect of the surface is to treat the body as a perfect solid, and emulate the surface effects by enclosing it in a 2D elastic membrane\cite{Gurtin}. Following this approach, it has been shown that surface roughness leads to a membrane enclosure which can have a negative effective Young's modulus\cite{Mohammadi}. For resonator structures, it was further shown that this decreases the resonant frequency in comparison to using the bulk value. As a result, the Young's modulus can be replaced by an effective one with a value below the bulk value.

While we have chosen to use the thickness as measured by the profilometer, $i.e.$ $t=20$~nm, when fitting the frequency tuning, it is also possible to fit the data assuming larger thicknesses, up to $t\approx 25$~nm. However, increasing the thickness requires using a higher effective Young's modulus to fit the measured data. From indentation measurements, we have an upper bound $E\le 180$~GPa which in practice limits the thickness that can be used in the fitting.

\section*{Mechanical model and resonant response}
This section details the mathematical procedure for finding the resonant frequencies of the Euler-buckled beam subject to an external (electrostatic) load. The fitting of the frequency tuning curves, including both geometric and electrostatic nonlinearities, were done by numerically solving the resulting equations for the stationary problem [Eqs.~(\ref{eq:buckl}) and (\ref{eq:tbar})] and then for the eigenfrequencies [Eq.~(\ref{eq:eig})].

\begin{figure}[t]
\begin{center}
\includegraphics*[width=0.9\linewidth]{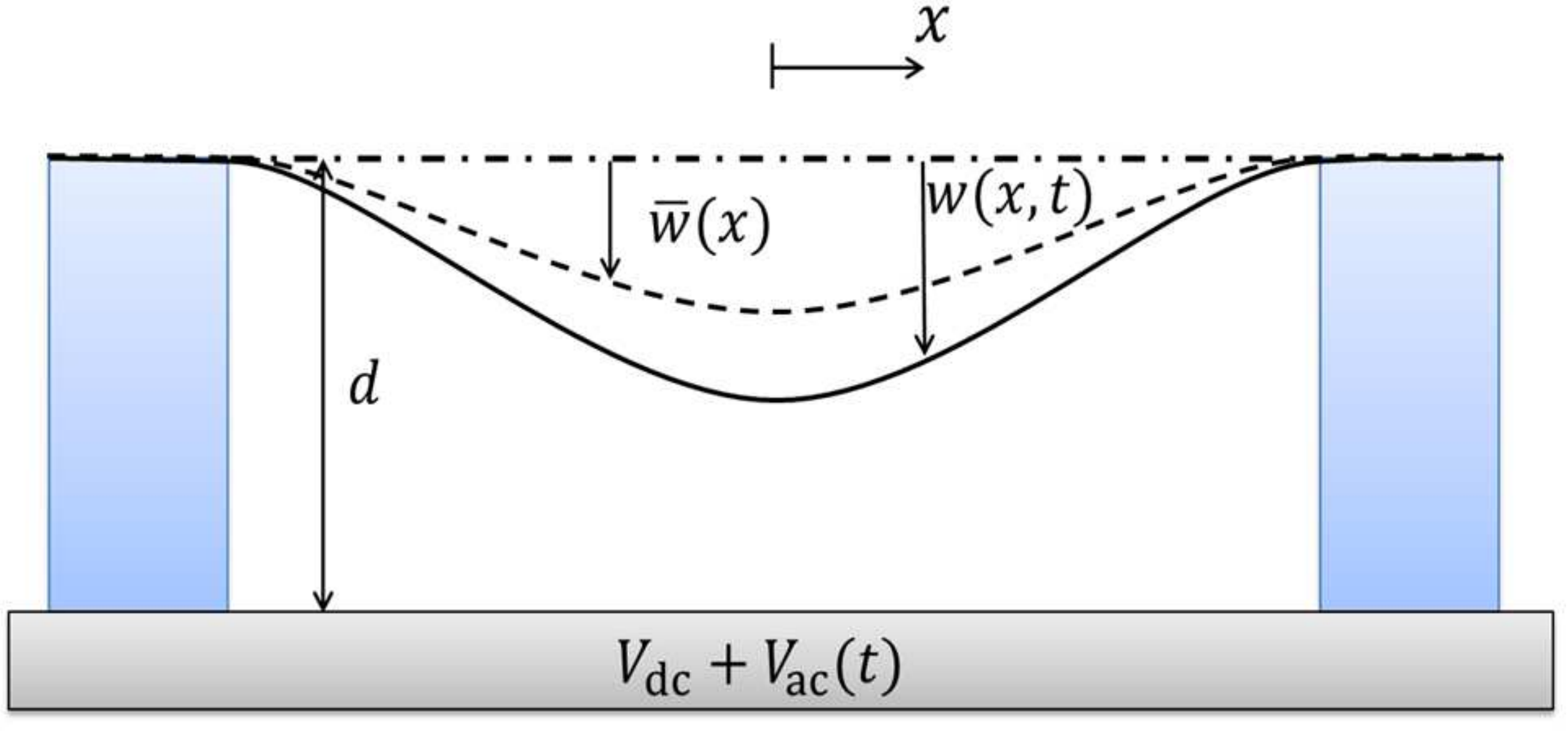}
\caption{Schematic sideview of a downward buckled DLC membrane of length $L$, downward deflection $w(x,t)=\overline{w}(x)+\delta w(x,t)$ subject to a bias voltage $V(t)=V_{\rm dc}+V_{\rm ac}(t).$\label{fig:modelfig}}
\end{center}
\end{figure}

To obtain the necessary equations, we model the suspended part of the DLC resonator as a wide uniform plate with the material parameters shown in Table~\ref{tab:parameters}. This makes the system essentially a 1D problem, namely that of the vibrations of a buckled beam as shown in Fig.~\ref{fig:modelfig}.

For a bias voltage $V\left(t\right){\rm =}V_{\rm dc}{\rm +}V_{\rm ac}{\rm (}t{\rm )}$,
the equation of motion using the Euler--Bernoulli beam theory is in the parallel-plate approximation\cite{Graff}
\begin{eqnarray}
&&\rho hW\partial^2_tw+\frac{EWh^3}{12}\partial^4_xw-\sigma Wh\partial^2_xw=\frac{\epsilon_0V^2W}{2(d-w)^2},\nonumber\\
&&\sigma=\sigma_0+\frac{EWh}{2L}\int_{-L/2}^{L/2} dx\, (\partial_xw)^2,\nonumber\\
&& w(\pm L/2)=\partial_xw(\pm L/2)=0,
\label{eq:EB}
\end{eqnarray}
where the small correction due to a finite Poisson's ratio $\nu$ has been neglected.
Rescaling Eq.~(\ref{eq:EB}), $i.e.$, putting it on dimensionless form
with characteristic length scale $L$ and timescale $\left({12\rho
    L^4}/{Eh^2}\right)^{1/2}$ gives the system of equations
\begin{eqnarray}
&&\left[\partial^2_\tau+\partial^4_x-T\partial^2_x\right]w=f,\nonumber\\
&&T=-T_0+\alpha \int_{-1/2}^{1/2}dx\, \left(\partial_xw\right)^2,\nonumber\\
&&w(\pm 1/2)=\partial_xw(\pm 1/2)=0.
\label{eq:nondim}
\end{eqnarray}
Here, $T=12\sigma L^2/Eh^2$, $T_0=-12\sigma_0L^2/{Eh^2}$,
$f\approx f_0/\left[1-\frac{w}{d}\right]^2$, with $f_0=6\epsilon_0L^3V^2_{\rm dc}/d^2Eh^3$ and $\alpha=6L^2/h^2$.

\subsection*{Static solution}
To find the response of the sheet we first seek the static solution. Hence, we set $w=\overline{w}(x)+\delta w(x,t)$ and consider the static problem
$[\partial^4_x-\overline{T}\partial^2_x]\overline{w}=f_0(1-\overline{w}/d)^{-2}=\overline{f}$, with $\overline{T}=-T_0+\alpha
\int dx\, (\partial_x\overline{w})^2$.

Depending on the sign of the resulting static tension $\overline{T}$ one finds the solutions
{\begin{eqnarray}
&&\overline{w}_-=-\frac{\overline{f}}{2\overline{T}}\left(x^2-\frac{1}{4}\right)\nonumber\\
&&+\frac{\overline{f}}{2|\overline{T}|^{3/2}}\frac{\cos\left(\sqrt{|\overline{T}|}x\right)-\cos
  \left(\frac{1}{2}\sqrt{|\overline{T}|}\right)}{\sin
  \left(\frac{1}{2}\sqrt{|\overline{T}|}\right)}
\label{eq:buckl}
\end{eqnarray}}
for $\overline{T}<0$ and
\begin{eqnarray}
&&\overline{w}_+=-\frac{\overline{f}}{2\overline{T}}\left(x^2-\frac{1}{4}\right)\nonumber\\
&&+\frac{\overline{f}}{2|\overline{T}|^{3/2}}\frac{\cosh\left(\sqrt{|\overline{T}|}x\right)-\cosh\left(\frac{1}{2}\sqrt{|\overline{T}|}\right)}{\sinh\left(\frac{1}{2}\sqrt{|\overline{T}|}\right)}
\end{eqnarray}
for $\overline{T}>0$.
The equilibrium tension is found from solving the equations
\begin{equation}
\overline{T}=-T_0+\alpha\frac{\overline{f}^2}{\overline{T}^2}\left(\frac{2}{\overline{T}}+\frac{1}{12}+\frac{1}{8\sqrt{|\overline{T}|}}\frac{\sqrt{|\overline{T}|}+3\sin\sqrt{|\overline{T}|}}{\sin^2\frac{1}{2}\sqrt{|\overline{T}|} }\right)
\label{eq:tbar}
\end{equation}
for $\overline{T}<0$ and
\[\overline{T}{\rm =-}T_0{\rm +}\alpha \frac{\overline{f}^2}{{\overline{T}}^{{\rm 2}}}\left(\frac{{\rm 2}}{\overline{T}}{\rm +}\frac{{\rm 1}}{{\rm 12}}{\rm -}\frac{{\rm 1}}{{\rm 8}\sqrt{\overline{T}}}{\rm \ }\frac{\sqrt{\overline{T}}{\rm +3}{\sinh  \sqrt{\overline{T}}\ }}{{{\sinh }^{{\rm 2}} \frac{{\rm 1}}{{\rm 2}}\sqrt{\overline{T}}\ }}\right){\rm ,\ \ }\overline{T}{\rm >0}.\]
Note that although Eq.~(\ref{eq:buckl}) seems to give zero static deflection when $\overline{f}\rightarrow 0$, the Euler buckling solutions for $\overline{f}=0$ are contained in the solution  Eq.~(\ref{eq:buckl}) due to the limiting behavior of Eq.~(\ref{eq:tbar}) when $\overline{f}\rightarrow 0$.

\begin{table}[t]
\caption{Parameters and quantities used in the modeling of the DLC membrane resonator.\label{tab:parameters}}
\vspace*{2mm}
\begin{tabular}{c|c|c}
Quantity/parameter & Symbol & Value or range \\
\hline\hline Width, & $W$ & \\
\hline Length & $L$ & 1 ${\rm \mu}$m \\
\hline Thickness & $h$ & 20 nm ($<$25 nm) \\
\hline Mass density & $\rho $ & 2000 kg/m${}^{3}$ \\
\hline Young's modulus$\ $ & $E$ & 140 GPa to 180 GPa \\
\hline Suspension height & $d$ & 205 nm \\
\hline Bias voltage & $V{\rm (}t{\rm )}$ & \\
\hline In-plane stress & $\sigma\left(t\right)$ &  \\
\hline Initial compressive stress & $-{\sigma }_0$ & $<$2.0~GPa \\
\hline Vacuum permittivity & $\epsilon_0$ & ${\rm 8.854\times }{{\rm 10}}^{{\rm -}{\rm 12}}{\rm \ }$F/m \\
\hline Vertical deflection & $w{\rm (}x,t{\rm )}$ & $\left({\rm -}\frac{L}{{\rm 2}}{\rm <}x{\rm <}\frac{L}{{\rm 2}}\right)$ \\
\hline
\end{tabular}
\end{table}

\subsection*{Small vibrations around equilibrium}
To find the resonant mode shape and frequency of the fundamental mode, we linearize Eq.~(\ref{eq:nondim}) around the equilibrium solution $\overline{w}$ to obtain the
eigenvalue problem
\[\left[{\partial }^{{\rm 2}}_t{\rm +}{\partial }^{{\rm 4}}_x{\rm -}\overline{T}{\partial }^{{\rm 2}}_x{\rm -}\frac{{\rm 2}\overline{f}}{d{\rm -}\overline{w}}\right]\delta w{\rm =}\delta T{\rm \ }{\partial }^{{\rm 2}}_x\overline{w},\]
where $\delta T{\rm =2}\alpha \int dx{\rm \ }\left({\partial
}_x\overline{w}\right)\left({\partial }_x\delta w\right){\rm
  =-2}\alpha \int dx{\rm \ }\delta w{\rm \ }{\partial }^{{\rm
    2}}_x\overline{w}$.
This eigenvalue problem has the general form
$$(\partial^4_x-T\partial^2_x-\lambda^2)u=-h(x)\int dx' h(x')u(x')$$ with ${\lambda }^{{\rm
    2}}{\rm =}{\omega }^{{\rm 2}}{\rm +}{{\rm
    2}\overline{f}}/{(d-\overline{w})}$
and
$$h(x)=\left\{ \begin{array}{c}
\sqrt{2\alpha}\frac{\overline{f}}{\left|\overline{T}\right|}\left(1-\frac{\sqrt{\left|\overline{T}\right|}}{2}
\frac{\cos\sqrt{\left|\overline{T}\right|}x}{{\sin
    \frac{1}{2}\sqrt{\left|\overline{T}\right|}\ }}\right){\rm
  ,\ \ }\overline{T}<0 \\ \sqrt{{\rm 2}\alpha
}\frac{\overline{f}}{\left|\overline{T}\right|}\left(-1+\frac{\sqrt{\left|\overline{T}\right|}}{{\rm 2}}{\rm
  \ }\frac{{\cosh \sqrt{\left|\overline{T}\right|}x\ }}{{\sinh
    \frac{1}{
        2}\sqrt{\left|\overline{T}\right|}\ }}\right),\,\,\overline{T}>0. \end{array} \right.$$

Following Ref.~\cite{Nayfeh} we set
$u=u_0+Ah(x)$ and note that $\left({\partial
}^4_x-\overline{T}{\partial }^2_x\right)h\left(x\right)=0$. This implies that the
solution to the problem is found by solving the simultaneous equations
\[\left({\partial }^{{\rm 4}}_x{\rm -}\overline{T}{\partial }^{{\rm 2}}_x{\rm -}{\lambda }^{{\rm 2}}\right)u_0{\rm =0,\ \ }A{\lambda }^{{\rm 2}}{\rm =}\int dx{\rm \ }h\left(x\right)\left[u_0\left(x\right){\rm +}Ah\left(x\right)\right].\]
Below we will restrict attention to the fundamental mode examined in the main part of the paper.

\subsection*{Fundamental mode resonant frequency, compressive stress}
For $\overline{T}<0$ we make the following Ansatz for the
fundamental mode
\[u_0{\rm =}B{\cos  \beta x\ }{\rm +}C{\cosh  \gamma x\ }\]
leading to the secular equations ${\beta }^4-\left|\overline{T}\right|{\beta }^2-{\lambda }^2=0$ and ${\gamma }^4+\left|\overline{T}\right|{\gamma }^2-{\lambda }^2=0.$ To satisfy the boundary conditions one must further fulfill the relations
\[Ah\left({\rm 1/2}\right){\rm +}B{\cos  \beta {\rm /2}\ }{\rm +}C{\cosh  \gamma {\rm /2}\, }=0\]
\[Ah^{{\rm '}}\left({\rm 1/2}\right){\rm -}B\beta {\sin  \beta {\rm /2}\ }{\rm +}C\gamma {\sinh  \gamma {\rm /2}\ }{\rm =0.}\]
As we are only interested in the spectrum, we are free to choose the normalization such that $A=1.$ Hence, to find the frequency of the fundamental mode under compressive stress one should solve the equation
\begin{equation}
{\lambda }^{{\rm 2}}{\rm =}\int dx{\rm \ }h{\rm (}x{\rm )[}B{\cos  \left(\beta x\right)\ }{\rm +}C{{\rm cosh\ } \gamma x\ }{\rm +}h{\rm (}x{\rm )]}.
\label{eq:eig}
\end{equation}
Here $\beta$, $\gamma$, $B$ and $C$ depend on ${\lambda }^2$ according to
\begin{equation}
\beta {\rm =}\sqrt{\sqrt{{\overline{T}}^{{\rm 2}}{\rm /4+}{\lambda }^{{\rm 2}}}{\rm -}\overline{T}{\rm /2\ }}{\rm ,\ \ }\gamma {\rm =}\sqrt{\sqrt{{{\overline{T}}^{{\rm 2}}}/{{\rm 4}}{\rm +}{\lambda }^{{\rm 2}}}{\rm +}{\overline{T}}/{{\rm 2}}{\rm \ }},
\end{equation}
\begin{equation}
B{\rm =-}\frac{\gamma h\left({{\rm 1}}/{{\rm 2}}\right){\sinh  {\gamma }/{{\rm 2}}\ }{\rm -}h^{{\rm '}}\left({{\rm 1}}/{{\rm 2}}\right){\cosh  {\gamma {\rm \ }}/{{\rm 2}}\ }}{\beta {\cosh  {\gamma }/{{\rm 2}}\ }{\sin  {\beta }/{{\rm 2}}\ }{\rm +}\gamma {\sinh  {\gamma }/{{\rm 2}}\ }{\cos  {\beta }/{{\rm 2}}\ }}{\rm ,\ }
\end{equation}
\begin{equation}
C{\rm =-}\frac{\beta h\left({{\rm 1}}/{{\rm 2}}\right){\sin  {\beta }/{{\rm 2}}\ }{\rm +}h^{{\rm '}}\left({{\rm 1}}/{{\rm 2}}\right){\cos  {\beta }/{{\rm 2}}\ }}{\beta {\cosh  {\gamma }/{{\rm 2}}\ }{\sin  {\beta }/{{\rm 2}}\ }{\rm +}\gamma {\sinh  {\gamma }/{{\rm 2}}\ }{\cos  {\beta }/{{\rm 2}}\ }}.
\end{equation}
The mode shapes and resonant frequencies can now be found by solving the eigenvalue equation [Eq.~(\ref{eq:eig})] for $\lambda^2$. While some progress can be made analytically by carrying out the integrals,  a numerical solution is needed to find the spectrum.\\

\subsubsection*{Fundamental mode, tensile stress}
For completeness we also briefly comment on the solution for tensile stress $\overline{T}>0$.
In this case the same Ansatz can
be used for the fundamental mode shape, $i.e.$, $u_0{\rm =}B{\cos \beta
  x\ }{\rm +}C{\cosh \gamma x\ }$. However, the equations for
wavenumbers become reversed, and we now get ${\beta
}^4+\left|\overline{T}\right|{\beta }^2-{\lambda }^2=0$ and ${\gamma
}^4-\left|\overline{T}\right|{\gamma }^2-{\lambda }^2=0.$ The formulas
for coefficients $B$ and $C$ remain unchanged.\\



\begin{thebibliography}{1}
\bibitem{pvd} More specifically, we refer to fabrication technique as filtered cathodic arc plasma accelerator deposition (FCAPAD). Alternatively, it can also be called filtered cathodic vacuum arc (FCVA).
\bibitem{prestress} The compressive stress was determined by measuring the curvature of a DLC-coated silicon sample with a profilometer and calculating the stress by using the Stoney equation. Typically, the same PVD process at room temperature yields a compressive stress of approximately 4 GPa, so there is a significant decrease in the stress due to the in-situ annealing at 500~${}^\circ$C.
\end{thebibliography}
\end{document}